\documentclass[aps,prb,preprint,endfloats*,superscriptaddress]{revtex4-1}

\usepackage{amsmath}
\usepackage{graphicx}
\usepackage{epsfig}

\newcommand{\lSAW}{\lambda_\mathrm{SAW}}	
\newcommand{\kSAW}{k_\mathrm{SAW}}			
\newcommand{\fSAW}{f_\mathrm{SAW}}			

\newcommand{\mycomment}[1]{}						

\begin{document}


\title{Optical phonon modulation in semiconductors by surface acoustic waves}


\author{F. Iikawa}
\affiliation{Paul-Drude-Institut f\"ur Festk\"orperelektronik, Hausvogteiplatz 5-7, 10117 Berlin, Germany}
\affiliation{Instituto de F\'isica ``Gleb Wataghin'', Unicamp, 13083-859, Campinas-SP, Brazil}
\author{A. Hern\'andez-M\'inguez}
\affiliation{Paul-Drude-Institut f\"ur Festk\"orperelektronik, Hausvogteiplatz 5-7, 10117 Berlin, Germany}
\author{M. Ramsteiner}
\affiliation{Paul-Drude-Institut f\"ur Festk\"orperelektronik, Hausvogteiplatz 5-7, 10117 Berlin, Germany}
\author{ P. V. Santos}
\affiliation{Paul-Drude-Institut f\"ur Festk\"orperelektronik, Hausvogteiplatz 5-7, 10117 Berlin, Germany}
\email[]{santos@pdi-berlin.de}


\date{\today}

\begin{abstract}
We investigate the  modulation of  optical phonons in semiconductor crystal  by surface 
acoustic wave (SAW) propagating on the crystal surface. The SAW  fields  induce changes on the order of 10\textsuperscript{-3}  in the average Raman scattering intensity by optical phonons in Si and GaN crystals. The SAW-induced modifications in the Raman cross-section are dominated by the modulation of the optical phonon energy by the SAW strain field. In addition to this local contribution, the experiments give evidence for a weaker and non-local contribution arising from the spatial variation of the SAW strain field. The latter is attributed to the activation of optical modes with large wave vectors and, therefore, lower energies. The experimental results, which are well described by  theoretical models for the two contributions, prove that optical phonons can be manipulated by SAWs with $\mu$m wavelengths. 
\end{abstract}


\maketitle


\section{Introduction}
\label{Introduction}

Following the successful developments in photonics during recent years, phononics is presently attracting considerable attention and is establishing itself as a promising area for future science and technological 
applications. The conventional way of controlling phonons takes advantage of size confinement.\cite{Richter81a,Campbell_SSC58_739_86,Woggon_PRB54_1506_96} The advances in epitaxial crystal growth have made it possible to fabricate nanostructures with atomic dimensional control. These have been used to tailor the propagation properties of acoustic\cite{NSCGW79a,Jusserand_PRB30_6245_84} and optical\cite{Sood_PRL54_2111_85} phonons, as well as their  interaction with  optical fields.~\cite{Fainstein95a, 
Lanzillotti-Kimura_PRB75_24301_07}

One interesting question is whether optical phonons can also be controlled in a dynamic way by an external stimulus. To our knowledge, only few approaches for active control of optical phonons have so far been proposed. One example is the  modulation of the optical modes in
strongly piezoelectric materials by an acoustic wave:\cite{Vtyurin_F170_181_95} here, the anisotropic strain field created by the sub-MHz bulk acoustic wave has been shown to reduce the crystal symmetry, thus introducing changes in the Raman scattering from optical phonons. 

In this paper,  we demonstrate that surface acoustic waves (SAWs) provide a convenient tool for  the  
dynamic manipulation of optical phonons in semiconducting materials.
The SAWs used here are elastic surface vibrations in the GHz range, which are electrically excited on the semiconductor surface using interdigital transducers. 
We demonstrate that the Raman spectrum from optical phonons in Si and GaN crystals becomes modified in the presence of a SAW with a wavelength $\lSAW$ of a few $\mu$m. For both materials, the changes in the Raman cross-section are on the order of 0.1\%. A systematic study shows that the  SAW-induced modifications in the Raman intensity and line shape are due to two mechanisms. The first is the periodic modulation of the optical phonon energy by the SAW strain field. Since the energy shifts are much smaller than the Raman line width, this mechanism leads to an apparent broadening of the Raman line recorded in continuous-wave (CW) experiments. In addition to the previous contribution, which is of local nature, the Raman experiments give evidence for a weaker and non-local contribution arising from the spatial variation of the SAW fields. The latter induces scattering events, which activate  modes with large wave vectors and, consequently, lower energies of the convex optical phonon dispersion of the crystal. 
The experimental Raman profiles are well reproduced by quantitative models for the two contributions. The results prove, therefore, the feasibility of  optical phonons manipulation by SAWs with $\mu$m wavelengths.

In the remaining of this manuscript, we first describe the sample preparation and the experimental setup to record the changes in the Raman spectrum by SAWs. We then present experimental results obtained for both Si and GaN crystals. In a subsequent discussion section, we develop a theoretical model for the acoustic modulation and compare it to the experimental findings. The last section summarizes the main conclusions of this work. 

\section{Experimental details}

The studies were carried out on electronic quality (001) Si and (0001) GaN substrates containing interdigital transducers (IDTs, see Fig.~1) for the excitation of SAWs fabricated by optical lithography. The split-finger IDTs were designed to launch SAWs with an acoustic wavelength $\lSAW=5.6~\mu$m  along a $\langle110\rangle$ ($\langle11\bar20\rangle$) surface direction of the  Si (GaN) wafer, respectively. In the case of Si, SAW excitation was enabled by coating the substrates with a 300 nm-thick piezoelectric ZnO film produced by magnetron sputtering prior to the IDT deposition. The corresponding IDT resonance 
frequencies at room temperature are 746 MHz and 1390 MHz for Si and GaN, respectively.

The Raman scattering studies were performed at room temperature in the  backscattering configuration using a confocal Horiba-LabRam system. The incident light from a 473 nm solid-state laser was focused by a 50$\times$ long working distance objective with a numerical aperture (NA) of 0.55 on a small spot (diameter of approx. 1~$\mu$m, which is much smaller than $\lSAW$) on the sample surface. The backscattered light was collected by the same objective, spectrally analyzed by a monochromator with a dispersion grating with  2400 grooves/mm, and detected using a liquid-nitrogen cooled charge coupled device (CCD).

The changes in the Raman intensity induced by SAW are very small (on the order of $10^{-3}$).  Special precautions are thus required to discriminate the SAW-induced changes from variations arising from fluctuations in laser fluency and  temperature  (caused, for instance, by 
heating due to the application of radio-frequency ({\it rf}) nominal peak power of typically 23~dBm to the IDT). For that purpose, we used  a modulation detection scheme similar to the ones described in Ref.~\onlinecite{Vtyurin_F170_181_95}. In this approach, both the excitation laser and the {\it rf}-power for SAW generation are on/off amplitude modulated at a frequency of 617~Hz. Raman spectra $I_R(\omega)$ were then recorded with the light and {\it rf}-modulation in phase (denoted as \textit{SAW-on} condition) and out-of-phase (\textit{SAW-off} condition) with an integration time per measurement $t_m\sim 1$~min. The following periodic recording sequence with four steps per cycle $i$ was used: $I^{i}_\mathrm{SAW,on}(\omega)$, $I^{i+1}_\mathrm{SAW,off}(\omega)$,  $I^{i+2}_\mathrm{SAW,off}(\omega)$, $I^{i+3}_\mathrm{SAW,on}(\omega)$. This sequence is then repeated  over $N\approx 60$  cycles leading to a total accumulation time of typically four hours. From the individual spectra we have determined the average Raman intensities in the SAW-on and SAW-off conditions,  
$I_\mathrm{R,on} (\omega)=\frac{1}{4 N}\sum_{i=1}^{N}( I^{i}_\mathrm{SAW,on} + I^{i+3}_\mathrm{SAW,on}  )$  
and
$I_\mathrm{R,off} (\omega)=\frac{1}{4 N}\sum_{i=1}^{N}( I^{i+1}_\mathrm{SAW,off} + I^{i+2}_\mathrm{SAW,on}  )$.  
From these two spectra, we then directly determined the total 
$I_\mathrm{R} (\omega)=I_\mathrm{R,on} (\omega) + I_\mathrm{R,off} (\omega)$
and the differential Raman spectrum defined as 
$\Delta I_\mathrm{R} (\omega)=I_\mathrm{R,on} (\omega) - I_\mathrm{R,off} (\omega)$.  

The modulation period $4t_m$ has been chosen to be short in comparison with the typical times for temperature and laser fluency fluctuations. As a result,  $\Delta I_{R}(\omega)$ reflects the changes introduced by the SAW corrected for random fluctuations in temperature and laser fluency, which are linear in time. This data acquisition scheme, which proved to be   essential for obtaining reliable data over the total integration times, was automatically  controlled by a  computer program.

\begin{figure}[tbhp]
\begin{center}
\includegraphics[width=.7\columnwidth,trim=70 350 100 280 , angle=0, clip]{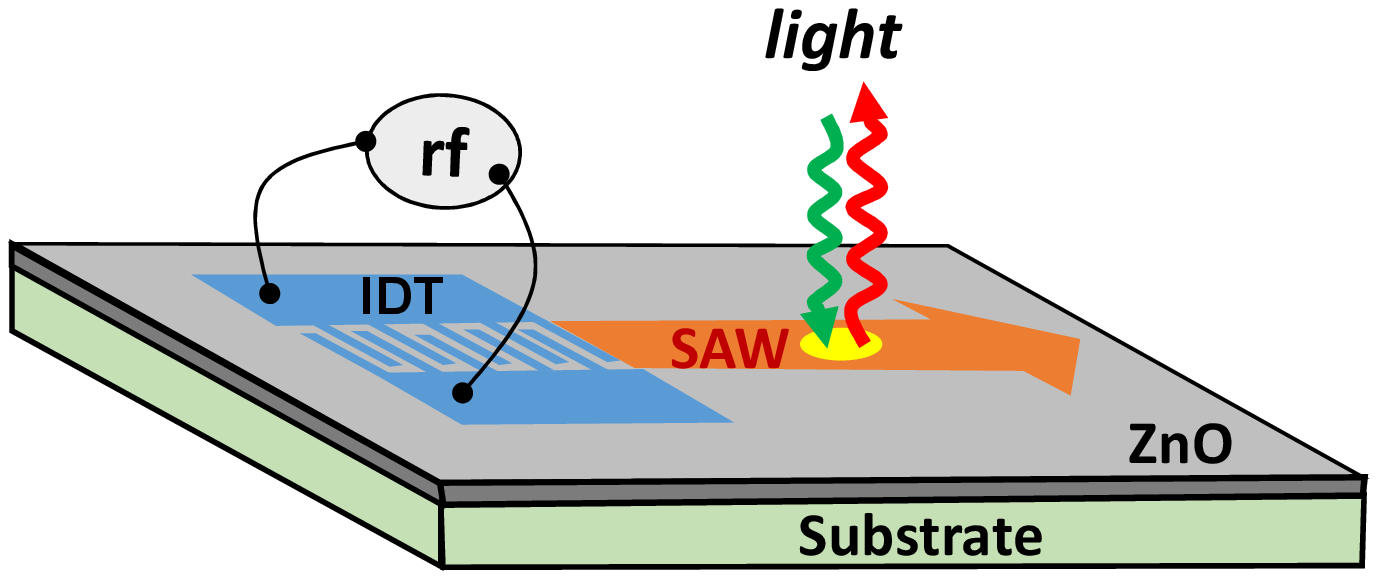}
\end{center}
\caption{Schematic diagram of a SAW delay line with interdigital transducers (IDT, only one IDT is shown) oriented along  a $\langle$110$\rangle$ surface direction of a (001) silicon substrate coated with a piezoelectric ZnO thin film. The  ZnO layer is absent in the GaN delay lines. The SAW propagation and incident and scattered light  direction are shown.}
\label{Fig1}
\end{figure}

\section{Results}

\begin{figure*}[tbhp]
\begin{center}
\includegraphics[width=1.0\textwidth, angle=0, keepaspectratio=true,trim=100 330 50 250,clip]{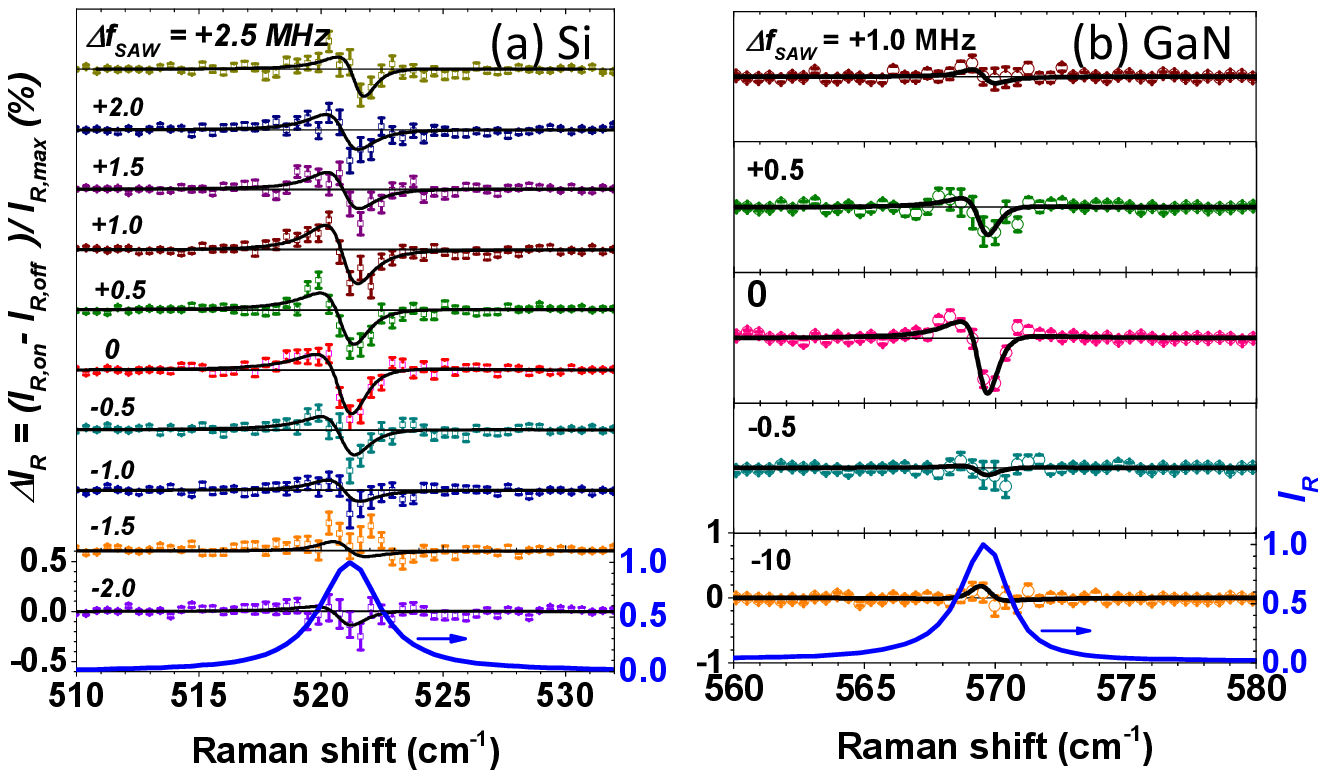}
\end{center}
\caption{Normalized total $I_{R} (\omega ) $ and differential $\Delta I_{R}(\omega)$ Raman spectra for (a) Si and (b) GaN recorded for different {\it rf}-excitation frequencies relative to the IDT resonance frequency 
$\fSAW$. Each spectrum is normalized to its maximum  value. The resonance frequency $f_0$ is equal to 746 MHz and 1.39 GHz for Si and GaN, respectively, and the resonance band approximately  1 MHz wide (cf. Fig. 3). The error bars were obtained from the statistical analysis of the Raman spectra ensemble (see text for detail). The solid lines obtained from the fits of Eq.~(\ref{EqFit}) to the experimental data are also shown.
\label{FigSpectra}
}
\end{figure*}

The bottom curves in Figs.~\ref{FigSpectra}(a) and \ref{FigSpectra}(c) display normalized backscattering Raman spectra for Si and GaN recorded in the absence of a SAW, respectively. The Si Raman spectrum  is characterized by a single peak  at $521.2$~cm$^{-1}$ associated with longitudinal optical (LO) phonons and possessing with a full width at half maximum (FWHM) of $3.0$~cm$^{-1}$. This line width, which is significantly larger than the spectral resolution of $0.5$~cm$^{-1}$,  attests to the high crystal quality of sample. The investigated GaN line at $570.0$~cm$^{-1}$ is associated with the excitation of the $E_{2}^{h}$  mode.~\cite{Siegle_PRB55_7000_97,Davydov_PRB58_12899_98} GaN has also a second Raman active $A_{1} (LO) $  mode at 735.5~cm\textsuperscript{-1}: its intensity, however, is lower than the one for the $E_{2}^{h} $  mode. Due to the worse signal/noise ratio, SAW modulation studies have not been carried out for the 735.5~cm\textsuperscript{-1} mode.

The additional curves in the panes of Fig.~\ref{FigSpectra} show differential Raman spectra $\Delta I_{R}(\omega)$ obtained using the modulation technique described above for different SAW frequency shifts relative to the IDT resonance frequency $\fSAW$. In each panel, the spectra have been normalized to the maximum value $I_\mathrm{R,max}$ of the $I_\mathrm{R}(\omega)$ spectrum recorded at $\fSAW$. The maximum modulation amplitude reaches $0.5$ and $0.8$~\% of $I_\mathrm{R,max}$ for Si and GaN, respectively, and reduces when the excitation frequency shifts away from $\fSAW$.

\begin{figure*}[tbhp]
\begin{center}
\includegraphics[width=1.0\textwidth, angle=0, trim=0 270 0 200,clip]{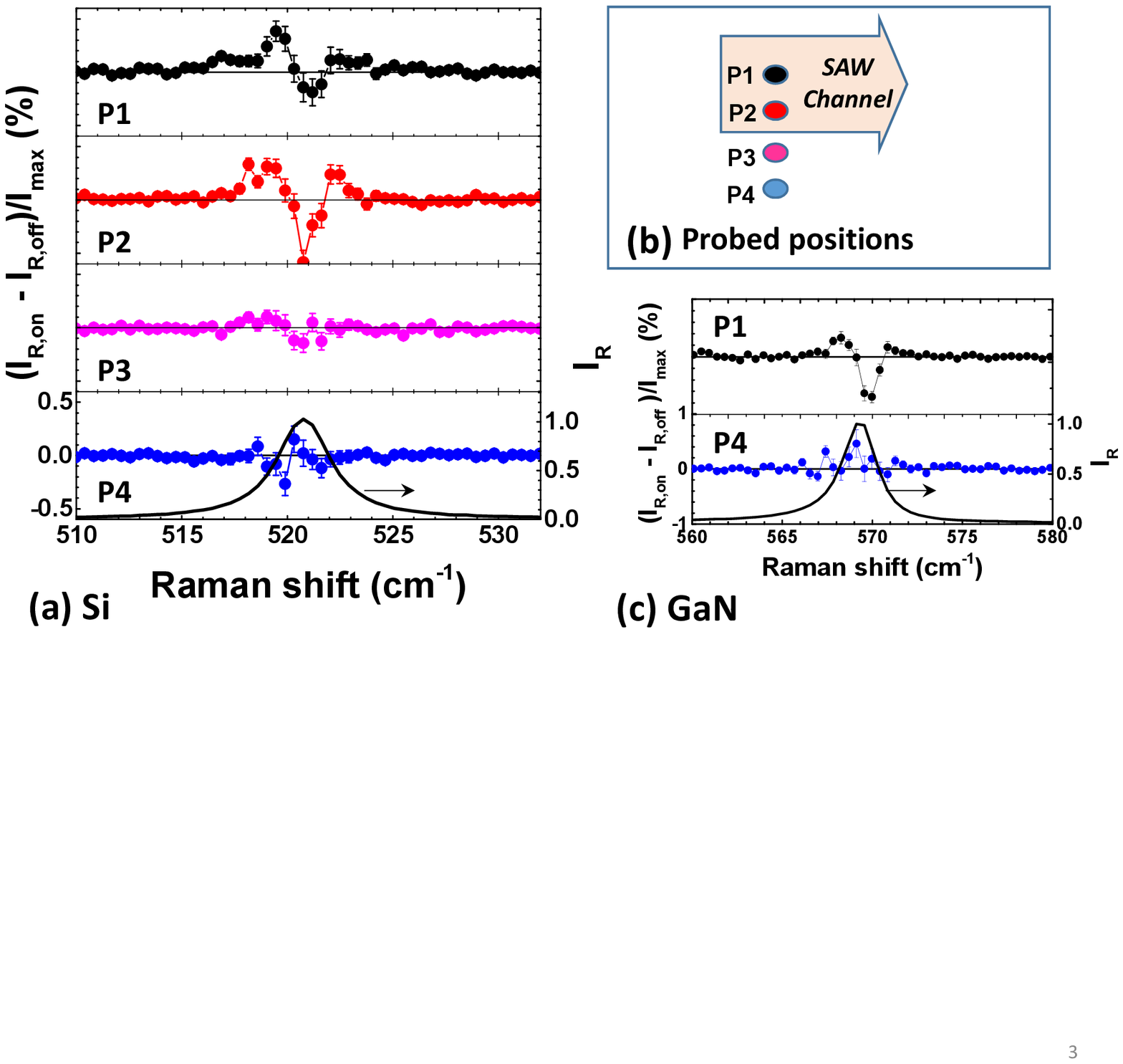}
\end{center}
\caption{Differential Raman spectra for (a) Si and (c) GaN recorded at the positions $P_1$-$P_4$ indicated in (b). $P_1$ and $P_2$ are within and $P_3$ and $P_4$ outside the 100-$\mu$m wide SAW beam (adjacent points are separated by 50~$\mu$m). 
\label{FigPos}
}
\end{figure*}

In order to demonstrate that the small changes in the Raman cross-section are really due to the SAW fields, we have recorded differential  Raman spectra by placing the laser spot at different positions on the sample surface.  Figures~\ref{FigPos}(a) and \ref{FigPos}(c)  compare results for Si and GaN, respectively, obtained on the positions $P_1$-$P_4$ indicated in Fig.~\ref{FigPos}(b). $P_3$ and $P_4$ are within and $P_1$ and $P_2$ outside the 100-$\mu$m wide SAW beam. All measurement spots are located 50~$\mu$m away for the IDT and separated by  50~$\mu$m.  
Note that a clear differential signal is only observed for the positions within the SAW beams. Furthermore, the temperature at all spots due to SAW excitation should be approximately the same since their separation is much smaller than the dimensions of the IDT. The different line shape and the small amplitude of the  signals recorded at $P_3$ and $P_4$ thus prove that the differential technique described in the previous section effectively corrects for temperature variations.

The differential Raman spectra  $\Delta I_\mathrm{R}(\omega)$ are characterized by a pronounced negative dip at the central Raman frequency with asymmetric flanks at  $\fSAW$, which indicates a preferential transfer of oscillator strength towards lower Raman shifts. In order to extract quantitative information from the differential spectra in Fig.~\ref{FigSpectra}(a) and 2(b), we fit each $\Delta I_{R}(\omega ) $  spectrum to the function:~\cite{Aspnes73a}

\begin{equation}        
f(\omega )={\rm Re} \left\{ \frac {Ae^{ \imath\theta }}{\left(\omega -\omega _{o} +i\gamma \right)^2} \right\},
\label{EqFit}
\end{equation}

\noindent where $A$, $\theta$, $\omega_0$ and $\gamma$ denote the amplitude, phase, resonance frequency, and spectral broadening, respectively.  This function is similar to the one used to analyze the line shape of modulation spectroscopy signals, including electro-reflectance and low frequency (hundreds of Hz) piezo-reflectance line shape.\cite{Gavini_PRB1_672_70}

\begin{figure*}[tbhp]
\begin{center}
\includegraphics[width=1.0\textwidth, angle=0, trim=0 330 20 290,clip]{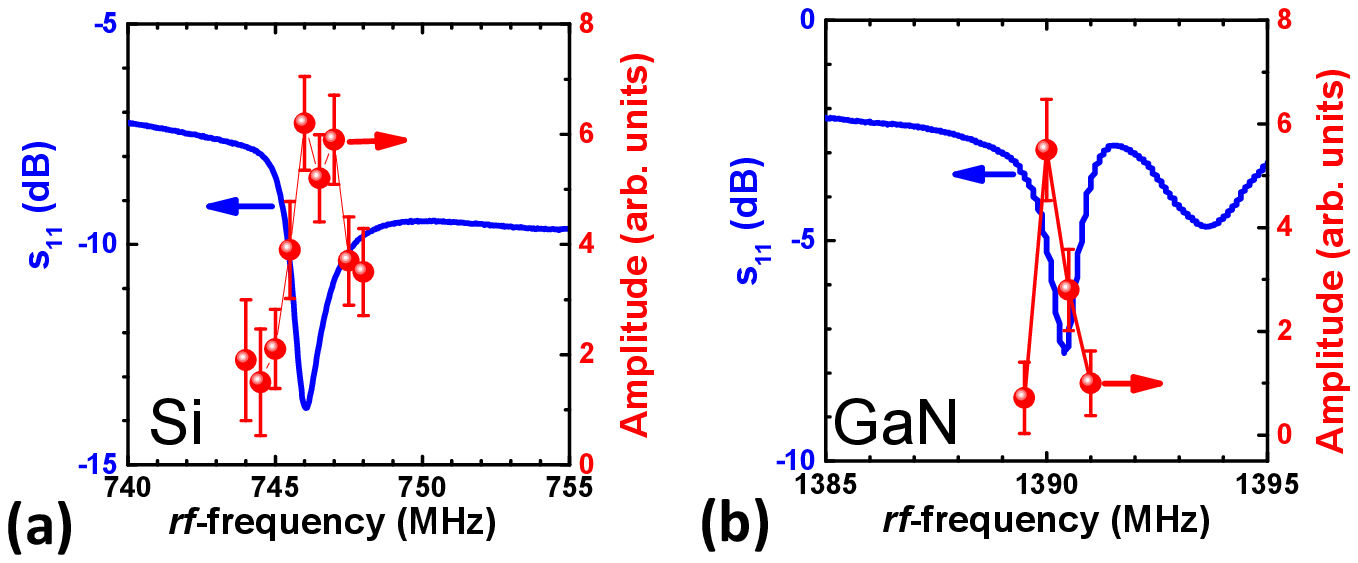}
\end{center}
\caption{The {\it rf}-reflectivity of the IDT (corresponding to the scattering parameter $s_{11}$, solid line)  and fitting amplitudes A (symbols) extracted from the curves of Fig.~\ref{FigSpectra} to Eq.~(\ref{EqFit}) (see text for detail). }
\label{FigS11}
\end{figure*}

The symbols in Figs.~\ref{FigS11}(a) and \ref{FigS11}(b) compare the 
amplitude \textit{A} obtained from the fits of Eq.~(\ref{EqFit}) to the experimental 
data of Fig.~\ref{FigSpectra} as a function of the {\it rf}-frequency applied 
to the IDT for Si and GaN, respectively. The solid line reproduces the frequency 
dependence of the {\it rf}-reflectivity of the IDT (corresponding to the {\it rf}-scattering parameter $s_{11}$) with the characteristic dip within the frequency band for SAW generation. The amplitude \textit{A 
} reaches a maximum within the resonance frequency range of the IDT. 
In the case of Si, while $\Delta I_{R}(\omega ) $ reduces considerably away from $\fSAW$, it 
does not vanish completely since the measured range of frequencies is not sufficiently far away from emission band of the IDT. 
Since the Raman measurements were carried out very close to the IDT (approx. 50~$\mu$m away), the residual signal may also be due to the excitation of other (non-SAW) acoustic modes by the IDT. Nevertheless, the results in Figs.~\ref{FigS11}(a) and \ref{FigS11}(b) confirm that the changes in the Raman spectra are due to the SAW strain field.

\section{Discussion}

The zone center optical modes in Si consist of one longitudinal (LO) and two transverse (TO) modes. In the absence of a SAW, these three modes are degenerate, but only the LO mode is Raman active in the backscattering configuration normal to the (001) surface. 
This mode propagates perpendicular to the surface with wave vector $q_z\hat e_z =\frac{4\pi \tilde n}{\lambda_L}\hat e_z $ and energy given by the phonon dispersion $\omega(q_z\hat e_z)$. In these expressions, $\tilde n$ is the refractive index and $\hat e_z $ the unitary vector along the surface normal, which corresponds to the $z$ direction.

The acoustic field introduces a periodic time and spatial modulation of the optical phonon energies as well as of the dielectric properties of the medium. Since the SAW frequency is much smaller than the optical frequencies, we can safely assume that the optical phonon frequencies adiabatically follow the acoustic modulation. The acoustic modulation affects the spectral shape of the differential Raman spectra in two different ways. First, the strain field periodically modulates the optical phonon energies. Since the energetic shifts are much smaller than the spectral width of the Raman line, the main effect in the time integrated Raman spectrum is to transfer oscillator strength from the center to the flanks of the line. This mechanism is expected to produce  a symmetric  differential Raman spectrum with a negative dip at the Raman central frequency and positive values at the flanks, in agreement with the line shape of Eq.~(\ref{EqFit}) for $\theta=0$. This lineshape  qualitatively reproduces the main features of the data in Fig.~\ref{FigSpectra}.

In addition to the symmetric contribution described above, the differential Raman spectra also show an asymmetric contribution yielding a stronger intensity at low Raman shifts. The spatially resolved plots of Fig.~\ref{FigPos} show that this contribution cannot explained by changes in temperature induced by SAW excitation. We attribute this contribution to phonon scattering processes arising from the spatial modulation of the refractive index and phonon energies by the SAW field. This non-local  mechanism  reduces  the crystal translation invariance symmetry and induces (i) a mixing of the TO and LO modes of the undisturbed crystal and (ii)  the coupling of dispersion modes $\omega({\bf q})$ with different ${\bf q}$'s.   Due to the negative curvature of the optical phonon dispersion, the last mechanism transfers  oscillator strength from the high to the low frequency flank of the Raman line, thus leading to an asymmetric contribution to $\Delta I_R(\omega)$. Such a process  has been invoked to account for the Raman line shape for nanocrystals. \cite{Richter81a,Campbell_SSC58_739_86} A similar behavior is also observed in Fig.~\ref{FigSpectra} and can be reproduced by setting $\theta=\pi/2$ in Eq.~(\ref{EqFit}).
 
The combination of the  mechanisms described above qualitatively reproduce the spectral shape of the $\Delta I_{R}(\omega)$ curves in Fig.~\ref{FigSpectra}. In the following sections, we proceed to a quantitative analysis of the impact of these mechanisms together with a comparison with the experimental results. For reasons which will become clear latter, we will concentrate on the results for the Si sample.

\begin{figure*}[tbhp]
\begin{center}
\includegraphics[width=1.0\textwidth, keepaspectratio=true, angle=0, trim=0 320 0 250,clip]{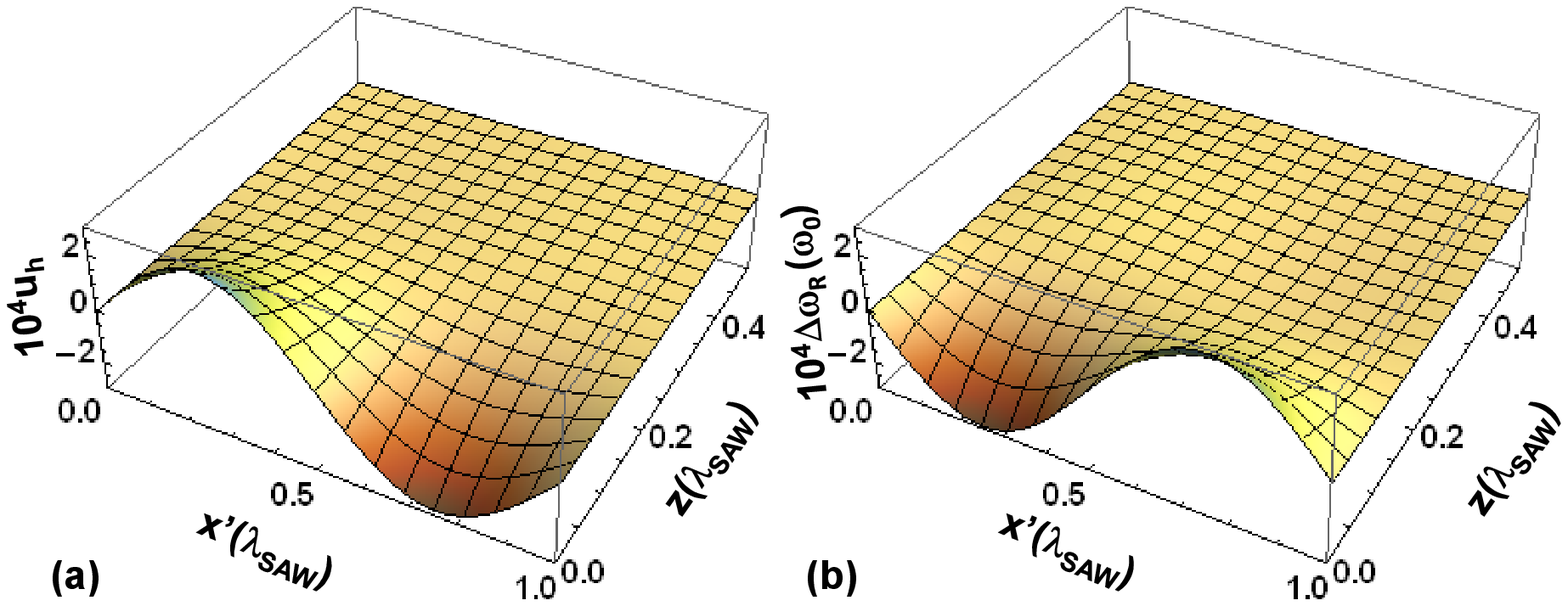}
\caption{Spatial distribution of the (a) hydrostatic strain $u_h=u_{x'x}'+u_{zz}$ and (b) local phonon frequency shifts $\Delta\omega_{R}$ induced by SAW with a wavelength $\lSAW=5.6~\mu$m propagating along the $x'=[110]$ direction of the ZnO/Si structure (see Fig.~\ref{Fig1}). Note that $z=0$ corresponds to the coordinate of the ZnO/Si interface.}
\label{FigA1}
\end{center}
\end{figure*}

\subsection{Acoustic field distribution}


The strain field $\varepsilon$ of a Rayleigh SAW propagating along the $x'$ direction of the surface contains three non-vanishing components and can be expressed (in the Voigt notation)  as  $\varepsilon=(u_{x'x'},0,u_{zz},0,u_{x'z},0)$. Here, $u_i$ ($i=x'||[110], y'||[\bar110], z||[001]$) are the displacement and $u_{ij}$ the strain  components.\cite{PVS156} 
The spatial distribution of the strain field can be obtained from an elastic continuum model for the structure following the procedure described in Ref.~\onlinecite{PVS156}. Calculations for the Si sample were carried out by taking into account the presence of the ZnO layer. Results for the hydrostatic component  $u_h=u_{x'x'}+u_{zz}$ induced by a SAW with wavelength $\lSAW=5.6~\mu$m and linear power density (e.g., the ratio between the acoustic power and the SAW beam width) of 76~W/m are shown in Fig.~\ref{FigA1}(a). In the plot, the ZnO/Si interface is at $z=0$ and the Si substrates fills the $z>0$ half plane. The calculation conditions correspond to the ones in Fig.~\ref{FigSpectra}(a) excitation of the IDT at its resonance frequency.  The absolute values for the strain components were obtained from the electric power coupled to the acoustic mode determined from the {\it rf}-reflection coefficient $s_{11}$ (cf. Fig.~\ref{FigS11}).

\subsection{Phonon frequencies}

In the presence of a strain field, the phonon frequencies $\omega$  of the zone center modes  can be written as $\omega^2 -\omega_0^2 = \lambda$, where $\omega_o$ is the unperturbed Raman frequency and $\lambda$ are the eigenvalues of the matrix~\cite{Cerdeira_PRB5_580_72}

\begin{widetext}
\begin{equation}
M_{ph}=
\begin{pmatrix}
{pu_{x'x'}}/{2}+q \left({u_{x'x'}}/{2}+u_{zz}\right) & 2 r u_{x'x'} & \sqrt{2} r u_{x'z} \\
 2 r u_{x'x'} & \frac{p u_{x'x'}}{2}+q \left({u_{x'x'}}/{2}+u_{zz}\right) & \sqrt{2} r u_{x'z} \\
 \sqrt{2} r u_{x'z} & \sqrt{2} r u_{x'z} & p u_{zz}+q u_{x'x'} \\
\end{pmatrix}.
\label{secularR}
\end{equation}
\end{widetext}

\noindent Here, $p$, $q$, and $r$ denote the phonon deformation potentials. In the following, we will consider again the case of Si and use the deformation potentials determined in Ref.~\onlinecite{Anastassakis_PRB41_7529_90} as  ${p}/{\omega_{0}^{2}}$ = -1.85,  ${q}/{\omega_{0}^{2}}$ = -2.30 and  ${r}/{\omega_{0}^{2}}$ = -0.71.
For the small strain levels achieved  under SAW modulation, the phonon frequency shifts are much smaller that the linewidth of the LO Raman line. Under this condition,  the Raman shift $\Delta\omega = \omega -\omega_0$ becomes $\Delta\omega \approx +\lambda/2\omega_0$. 


The solution of Eq.~(\ref{secularR}) for the SAW strain field distribution in Fig.~\ref{FigA1}(a) yields,  at each spatial/time position,  three eigenmodes $\lambda_i$ as well as three associated eigenvectors $\phi_i$. In order to evaluate the impact of these shifts on the Raman signal, we also have to take into account the coupling of the modes to the light field, which depends on the Raman selection rules. For that purpose, we will assume that the contribution of each eigenmode to Raman intensity is proportional to  its LO content, which can be obtained from the projection of  $p_{i,LO}=\left|\langle\phi_i|LO\rangle\right|^2$ of its displacement pattern of two $i$ = TO modes along the longitudinal direction. In this way, we obtain a map of the  effective local Raman shift given by $\Delta\omega_R=\sum_{i=1,3} {p_{i,LO} \lambda_i/2\omega_0}$. 

Figure~\ref{FigA1}(b) displays the spatial distribution of $\Delta\omega_R$ corresponding to the strain profile of Fig.~\ref{FigA1}(a). $\Delta\omega_R$ closely resembles the hydrostatic strain distribution with largest amplitudes  close to the $z=0$ (i.e., close to the ZnO/Si interface) and   positive frequencies shifts at the location of compressive strain (i.e., $u_h<0$). Similar calculations can also be carried out for GaN.

\begin{figure}[tbhp]
\begin{center}
\includegraphics[width=1.0\columnwidth, keepaspectratio=true, angle=0, trim=100 250 50 200,clip]{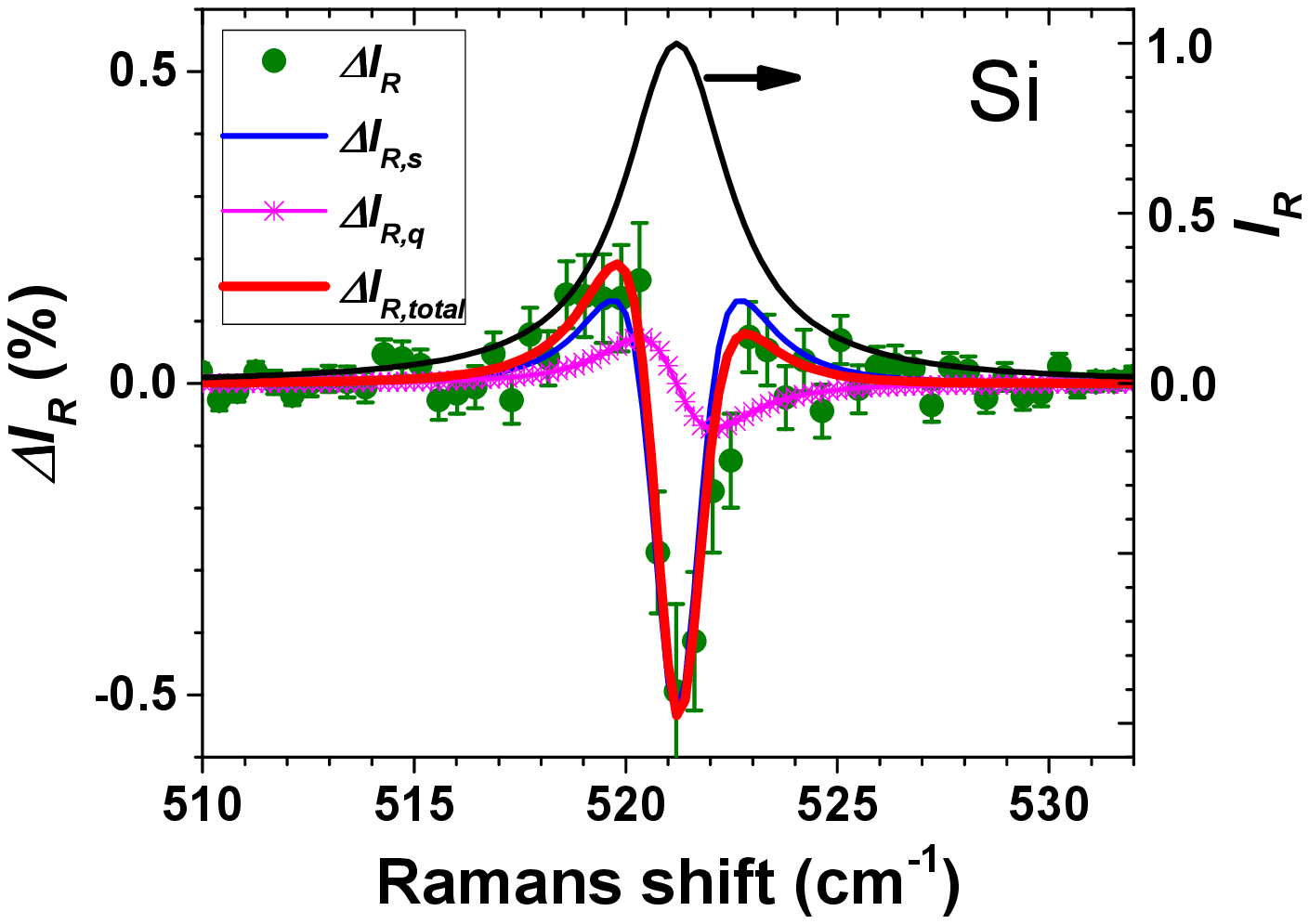}
\caption{Experimental and calculated differential Raman intensity of Si by taking into account only the strain-induced frequency modulation $\Delta I_{R,s}$, only the coupling between states with different $\bf q$ vectors  $\Delta I_{R,q}$, and total contribution  $\Delta I_{R,total}$. $I_{R}$ shows, for comparison, the unperturbed Raman spectrum.}
\label{FigA3}
\end{center}
\end{figure}

\subsection{Differential Raman spectrum}

In order to determine the modifications in the Raman spectrum induced by the acoustic field, we will assume that the optical phonons can be considered as particles with an effective mass given by the LO dispersion subjected to an effective potential given by  Fig.~\ref{FigA1}(b). The components of the Fourier transform of this potential profile act as interaction potentials for phonons with different wave vectors $\bf q$ and, consequently, different energies given by the wave vector dependence of the phonon dispersion $\omega({\bf q})$. In this way, one can  setup an  eigenvalue equation yielding the phonon modes under the SAW potential. The solution of this equation gives the phonon states of the perturbed system. 

Instead of solving the full eigenvalue problem, we examine two limiting approximations determined by the phonon coherence length $\ell_{ph}$. The latter is an important parameter governing the phonon properties in the modulated potential, since it determines the region over which the vibrational modes average the local potential variations displayed in Fig.~\ref{FigA1}(b). First, we consider the situation where $\ell_{ph}\ll\lSAW$, which is expected to hold here since $\lSAW$ is much larger than the lattice period. In this case, the Raman spectrum becomes proportional to the phonon density of states in the potential landscape of Fig.~\ref{FigA1}(a), which can be obtained by simply integrating the frequency-shifted contributions  over the scattering volume. For Si, the short optical penetration depth (which  corresponds to only 5\% of $\lSAW$) allows for  a further simplification, since the differential Raman signal can be obtained by a one-dimensional integration along $x'$ for $z=0$, i.e., 

\begin{eqnarray}
\Delta I_{R,s}(\omega) &=&\frac{1}{\lSAW}\int_0^{\lSAW} I_{R,0}\left[\omega, \omega_0+\Delta\omega_R(x',0)\right] dx'  \nonumber \\
& &-  I_{R,0}(\omega, \omega_0).
\label{EqRf}
\end{eqnarray}

\noindent Here  $I_{R,0}(\omega, \omega_0)$ denotes the spectral shape of the unperturbed Raman intensity with Raman line at $\omega_0$. The latter will be assumed to be a  Lorentzian function with line width corresponding to the measured value of 3.0~cm$^{-1}$. 

 The calculated $\Delta I_{R,s}$ displayed in  Fig.~\ref{FigA3} reproduces well the amplitude and spectral shape of the  main dip observed in the measured differential Raman spectrum  (symbols). The amplitude of the oscillation in the phonon frequency corresponds in this case to 0.16 cm$^{-1}$,  which is equal to approximately 10\% of half width of the Raman line.

While the previous model reproduces well most of the experimental features, it cannot account for the slightly asymmetric shape of the differential spectrum with respect to the center frequency, which is particularly strong for Si (cf. Fig.~\ref{FigSpectra}). The natural mechanism to account for the asymmetry is the coupling of modes  with different  ${\bf q}$'s induced by the acoustic modulation. Such a contribution is expected if the coherence length $\ell_{ph}$ becomes comparable to the spacial extent of the modulation potential created by the SAW. Under these conditions, the Fourier component ${\cal F}_{ph}({\Delta \bf q})$  will couple modes with wave vector differing by ${\Delta \bf q}$. Due to the parabolic shape of the phonon dispersion close to the zone center $\omega({\bf q}) = \omega(0) - {\bf ( q)^T a ( q)}$, where $\bf a$ describes the curvature of the LO dispersion,  the strongest contribution to the scattering comes from the large scattering vectors. According to  Fig.~\ref{FigA1}, these are found for ${\Delta \bf q}||z$ as a consequence of the strong decay of the acoustic field with $z$. 

We will estimate the effects of the ${ \Delta \bf q}$ coupling on the  differential intensity  $\Delta I_{R,q}$ using the phenomenological approach developed for nanocrystals by  Campbell and Fauchet.\cite{Campbell_SSC58_739_86} The latter assumes that the Raman spectrum involves contributions of $\bf q$ vectors around the $\Gamma$ point with a characteristic cut-off wave vector $q_{max}$, thus leading to a differential spectrum $\Delta I_{R,q}(\omega)$ given by

\begin{widetext}
\begin{equation}
\Delta I_{R,q}(\omega) =
\frac{1}{q_{max} \sqrt{2\pi}}
\int_{-\infty}^{\infty} I_{R,0}\left(\omega, \omega(q)\right) 
\exp{ \left[ -\left(   \frac{q}{q_{max} \sqrt{2} }\right)^2 \right]} dq -
 I_{R,0}(\omega(0)).
\label{Eqq}
\end{equation}
\end{widetext}

The differential intensity $\Delta I_{R,q}(\omega)$ calculated for $q_{max}=40~k_{SAW}$, which is shown by the crosses  in    Fig.~\ref{FigA3}, has an anti-symmetric, S-like line shape similar to the one observed in the experimental data. 
In fact, calculations for the  total differential spectrum including this contribution reproduce  well the experimental line shapes over the whole range of Raman shifts, as displayed by $\Delta I_\mathrm{R,Total}(\omega)$ in Fig.~\ref{FigA3}. 
One difficulty with the model is the large values for $q_{max}$ ($q_{max}\gg\kSAW=2\pi/\lSAW$), which implies the existence of non-vanishing Fourier components up to the 40$^{th}$ order. According to the phonon potential profiles (cf.~Fig.~\ref{FigA1}) the highest Fourier coefficients arise from the decay of the SAW amplitude with depth (rather than from the variation along the SAW propagation direction). The latter, however, leads to non-vanishing Fourier components only up to approx. $10~\kSAW$. The generation of phonons with large wave vectors can also proceed via multi scattering events, in particular under high SAW amplitudes. Further studies will be required to elucidate the origin of this strong non-local contribution.

Similar calculations can also be carried out for GaN. However, since the GaN substrate is transparent for the used laser line, one probes a sample depth larger than the SAW decay length, thus making a quantitative analysis more difficult. In addition, the experimental phonon dispersion for the E$^h_{2}$  mode is not well known,\cite{Ruf_PRL86_906_01} thus requiring additional fitting parameters.
Therefore, it needs a more complete calculation  that is out of scope of this work. Nevertheless, the asymmetric lineshape observed in GaN, shown in  Fig.~\ref{FigSpectra}b , similar to the Si, is  an indirect indication of  the negative dispersion of E$^h_2$ phonon mode  predicted by ab-initio calculations.\cite{Siegle_PRB55_7000_97,Bungaro_PRB61_6720_00,Ruf_PRL86_906_01}

\section{Conclusions}

We demonstrate that  surface acoustic waves can  modulate optical phonons in Si and GaN semiconductors. The SAW-induced variation in the Raman scattering cross-section, which is of the order of 10$^{-3}$, shows characteristic features attributed to two effects.  The main contribution is identified as the modulation of the phonon frequency by the SAW strain. The  second effect is the coupling of the phonon wave vectors induced by the non-homogeneous strain modulation. The latter is of non-local origin and associated with the short-range spatial variations of the SAW fields.  
The differential Raman intensity calculated using a model that includes both contributions is in good agreement with the experimental spectra. 

We thank V. Kaganer for discussions and for a careful reading of the manuscript. F. I. acknowledges the financial support from the AvH Foundation (Germany), Fapesp (Brazil), and CNPq (Brazil).



%

\end{document}